\documentclass[twocolumn,showpacs,preprintnumbers,amsmath,amssymb]{revtex4}
\usepackage{graphicx}% Include figure files
\usepackage{graphicx}% Include figure files
\usepackage{dcolumn}% Align table columns on decimal point
\usepackage{bm}% bold math
\usepackage{amsxtra}
\usepackage{latexsym}

\newcommand{\sline}{\mbox{\protect\rule[2pt]{0.4cm}{0.1pt}}}
\newcommand{\dashline}{\mbox{\protect\rule[2pt]{0.1cm}{0.1pt} \hspace{-4pt} 
    \protect\rule[2pt]{0.1cm}{0.1pt} \hspace{-4pt} \protect\rule[2pt]{0.1cm}{0.1pt}}}
\newcommand{\ddashline}{\mbox{$\cdot \cdot$ \hspace{-3pt}
    \protect\rule[2.1pt]{0.175cm}{0.6pt} \hspace{-3pt}
    $\cdot \cdot$}}

\newcommand{\circdott}{\mbox{\ensuremath{ \hspace{-1pt} \circ \hspace{-3.75pt} \cdot}}}

\begin{document}

\title{Cooperative sequential adsorption with nearest-neighbor
  exclusion\\ and next-nearest neighbor interaction}

\author{C.~J.~Neugebauer}
\email{cjn24@cam.ac.uk}
\affiliation{%
  Department of Chemistry, University of Cambridge,
  Cambridge, United Kingdom
}

\author{S.~N.~Taraskin}%
\affiliation{%
  St. Catharine's College and Department of Chemistry, University of Cambridge,
  Cambridge, United Kingdom
}%

\date{\today}

\begin{abstract}
A model for cooperative sequential adsorption that incorporates nearest-neighbor
exclusion and next-nearest neighbor interaction is presented.
It is analyzed for the case of one-dimensional dimer and
two-dimensional monomer adsorption.
Analytic solutions found for certain values 
of the interaction strength are used to investigate 
jamming coverage and temporal approach to jamming in the one-dimensional
case. 
In two dimensions, the series expansion of the coverage $\theta(t)$ is
presented and employed to provide estimates for the jamming coverage as a function
of interaction strength.
These estimates are supported by Monte Carlo simulation results.
\end{abstract}


% 05.20.-y classical statistical mechanics  
% 05.70.Ln Nonequilibrium and irreversible thermodynamics (see also 82.40.Bj
%          Oscillations, chaos, and bifurcations in physical chemistry and chemical
%          physics)       
% 68.43.De Statistical mechanics of adsorbates
% 68.43.-h 	Chemisorption/physisorption: adsorbates on surfaces

\maketitle

Surface adsorption phenomena are important in a great number of physical,
chemical and biological systems.
Equally large is the number of phenomena themselves that occur when 
particles/molecules are adsorbed to a surface.
The surface-adsorbate interactions can be broadly
classified into two categories, physisorption and chemisorption~\cite{gross_03}.
While the former is associated with electrostatic interactions,
van der Waals forces~\cite{ehrlich_64, autumn_00}, the latter is commonly
associated with the formation of 
chemical bonds.
Due to the strength of the chemical bond, chemisorption is often irreversible for
temperatures of interest.
An well-known example is 
the adsorption of water on Fe(001)~\cite{dwyer_77}.
A simple but effective model for such irreversible adsorption is the
\emph{random} or \emph{cooperative sequential adsorption}
(CSA)~\cite{evans_93}: particles are adsorbed randomly in a 
sequential manner without diffusing or desorbing.

For irreversible adsorption, a central quantity of interest is the
coverage, $\theta$, of
the final adsorbed monolayer attains, the jamming coverage,
$\theta_J$. 
Detailed knowledge of this jamming coverage might, for example, become
important for chemical sensing devices, such as 
micro-cantilevers~\cite{hu_01}, in order to distinguish between different species of
adsorbates. 
The jamming coverages have been estimated within several
approaches~\cite{evans_93}.  
However, important and significant effects due to interactions between
adsorbates on the jamming coverage have not been considered in detail 
and are the focus of this work. 

One of the simplest models for interaction between adsorbates is the 
\emph{nearest-neighbor exclusion} (NE) mechanism that causes an
adsorbed particle to block its nearest-neighbor binding
sites from adsorption~\cite{dwyer_77}. 
It has been shown~\cite{dickman_91_RSA} that for the NE process 
on a two-dimensional surface with
a square-lattice arrangement of binding sites the jamming coverage 
approaches a non-trivial value of $\theta_J = 0.3641$ which is lower
than that of the ideal limit, $\theta_0 = 0.5$. 
This is a consequence of the stochastic nature of the adsorption process
which results 
in the exponentially small probability of an ideally covered surface.

Here, we introduce, in addition to the NE, a short-range
adsorbate-adsorbate interaction 
which provides a more accurate and realistic description of adsorption
process. 
This interaction  affects  
the rate of adsorption of a next-nearest neighbor (NNN) binding site
and influences the jamming coverage. 
Such interactions might be caused by a variety of mechanisms.
If the adsorbates have an effective charge, then resulting electrostatic
forces would lead to a repulsive interaction: for example, hydrogen on Pd(100)
acquires a dipole moment due to charge transfer from the surface~\cite{gross_03}.
In this case, assuming that the adsorption rates are of Arrhenius
type and that each occupied NNN of an available binding site
contributes an equal amount to the binding energy $\varepsilon_{\text{int}}$,
the adsorption rate including the interaction with $n$ occupied NNN could be
modeled by $r_n = r_0 \exp \,(-n \, \varepsilon_{\text{int}} /T 
)$~\cite{pak_87} 
(where $r_0$ is a typical rate of adsorption and $T$ is the temperature).  
Attractive interaction might arise if an adsorbate 
(e.g.\ water on Pd(100) ~\cite{kampshoff_96, dong_98}) induces a local 
change in the surface structure~\cite{barnes_94}  
that increases the rate of adsorption at NNN sites.
Another mechanism that would lead to an effective attractive interaction could
involve precursor layer diffusion~\cite{king_74}: a gas particle might
become 
physisorbed even if it collides with an already adsorbed
particle~\cite{ehrlich_61}. 
In that case, it might either desorb or diffuse to the next available
site surrounding the adsorbed cluster for chemisorption.
 
Below, we use a linear approximation in $n$ for the 
adsorption rate, $r_n = 1 - n \epsilon$ ($r_0 =1$ by rescaling the
time). 
Such a linear dependence on $n$ arises naturally for the attractive interaction
of adsorption via the precursor state, as each occupied NNN should contribute
equally to the adsorption of the surrounded site~\cite{rodgers_97}.
It is clearly the first order approximation of the Arrhenius-type rate
presented above, 
which is valid for high temperatures or small interaction energies
$\varepsilon_{\text{int}}$ and also a possible assumption 
for the mechanism of morphology changes in the surface.  
It should be mentioned, that 
the effects of adsorption rates that depend on NNN occupation, especially 
on the island structure, have been previously considered~\cite{evans_nord_87},
albeit with a different choice of rates.   

The effects of the  NNN interaction on
adsorption can be investigated by means of rate
equations~\cite{evans_93} 
which  describe the evolution of the marginal probability density $P(G; t)$ of
finding a configuration $G$ of lattice sites empty at time $t$, irrespective
of the state (occupied or vacant) of the remaining sites of the lattice.
For the two-dimensional square lattice, the rate equations are~\cite{wang_00} 
\begin{eqnarray}
  \partial_t P(G;t) = - \sum_{i \in G} \sum_{n=0}^z r_n P (\{ G \cup
  D_i \}_n; t)~. 
  %\, \delta_{N_{nnn}(G,i), n}
  \label{eq:CSAmaster}
\end{eqnarray}
The only way that in an irreversible adsorption process the probability of
finding a set $G$ of sites can change is by adsorption at one of its binding
sites $i \in G$.
Due to NE, $i$ must have $z$ empty nearest neighbors.
Therefore if $i$ lies on the
boundary of $G$, this can only happen if $G$ is a subset of the larger set of
empty sites, i.e.\ $G \cup D_i$.
The subscript $n$ in $\{ G \cup  D_i \}_n$ refers to
 the additional interaction with the environment of site 
$i$ surrounded by $n$ occupied NNN. 
Using the fact that the marginal probability densities 
obey the following relation, 
 $P(G
\cup \{\sigma_j = 1\};t) + P(G \cup \{\sigma_j = 0\};t) = P(G; t)$ ($\sigma_i$
denotes occupation of site $i$) we can recast the RHS
of Eq.~(\ref{eq:CSAmaster}) completely in terms of
probability densities of configurations of empty sites.
For example, considering only the contribution to the rate equation due to
adsorption at the dotted site, Eq.~(\ref{eq:CSAmaster}) reads 
\begin{eqnarray}
\partial_t P \big(
%%%%
  \begin{array}{c@{\hspace{1pt}}c@{\hspace{1pt}}c}
    \vspace{-7.5pt} & \circ &     \\
    \vspace{-7.5pt} \circ & \circ & \circdott \\
    & \circ & 
  \end{array};
%%%
  t \big) &=& 
\ldots -(1-2 \epsilon) P \big(
%%%%
  \begin{array}{c@{\hspace{1pt}}c@{\hspace{1pt}}c@{\hspace{1pt}}c}
    \vspace{-7.5pt} & \circ &   \circ &         \\
    \vspace{-7.5pt} \circ & \circ & \circdott & \circ\\
             & \circ & \circ &  
  \end{array};
%%%
  t \big) \nonumber \\
  && - 2 \epsilon  P \big(
%%%
  \begin{array}{c@{\hspace{1pt}}c@{\hspace{1pt}}c@{\hspace{1pt}}c}
           \vspace{-7.5pt} & \circ &   \circ &          \\
    \vspace{-7.5pt} \circ & \circ & \circdott & \circ\\
             & \circ & \circ & \circ
  \end{array};
%%%
  t \big)~.    
\end{eqnarray}
Formally, we write such a rate equation as $\partial_t P(G;t) = - \mathcal{L}
P(G;t)$ where $\mathcal{L}$ is the operator that generates the configurations
$G'$ of empty sites that can produce $G$ by a single adsorption event. 

First, we analyze the situation in 1d.
Monomer adsorption with NE and NNN interaction is equivalent to dimer
adsorption with with nearest-neighbour interaction in 1d~\cite{evans_83,
  wolf_84}, which we will consider here. 
This model has been solved for general cooperative rates~\cite{gonzalez_73}.
However, the temporal approach to jamming, which is mainly of our
interest, is not readily available from such a solution. 
Therefore, we present here a different form of the solution that is
suitable for our purposes.

For this process, the following rate equations for finding a
stretch of $m$ vacant sites  can be written:
\begin{eqnarray}
  \partial_t P(1;t) &=& -2 (1 - 2 \epsilon) P(2;t) - 4 \epsilon P(3;t)
  \label{eq:1d_csa_1} 
\\
  \partial_t P(m; t)&=& -[m-1-2 \epsilon] P(m;t) - 2 P(m+1;t) \nonumber \\
  && - 2 \, \epsilon \, P(m+2;t)  \quad
  \mbox{for $m \geq 2$} \label{eq:1d_csa_n}
~. 
\end{eqnarray}
With the initial condition of an empty lattice,
Eq.~(\ref{eq:1d_csa_n}) is solved exactly by  
\begin{eqnarray}
  P(m;t) &=& \exp[ -(m-1-2 \epsilon) t - 2 (1 - e^{-t}) \nonumber
  \\ && - \epsilon (1 - e^{-2t})] 
  \label{eq:1d_csa_n_sol}. 
\end{eqnarray}
 for $m\ge 2$. 
The solution for the case  $m=1$ (see Eq.~(\ref{eq:1d_csa_1})) 
is given by 
\begin{eqnarray}
P(1;t)= 1-2 e^{-2-\epsilon} \left((1-2\epsilon) I_2^\epsilon (t) + 2 \epsilon
  I_3^\epsilon (t) \right)
 \label{eq:1d_csa_999} 
\end{eqnarray}
where 
\begin{eqnarray}
I_m^\epsilon (t) = e^{2 + \epsilon} \int_0^t P(m; t') \, \mathrm{d}t'~.
\end{eqnarray}
The probability $P(1;t)$ is particularly important for evaluation
of the critical coverage, 
$\theta_J = 1 - \lim_{t \to \infty} P(1;t)$. 
The integrals $I_m^\epsilon (t)$ can be evaluated analytically only
for some special cases. 
Namely,  
 by considering negative values for $\epsilon = - \alpha$ ($\alpha
 >0$), 
we can
write $I_m^\epsilon (t)$ in terms of a sum of lower incomplete
gamma functions $\gamma$~\cite{abramowitz+stegun},
\begin{eqnarray}
I_m^{-\alpha}(t) &=& \frac{1}{2} K_m^{\alpha} \sum_{k=0}^\infty \alpha^{k/2}
{m+2\,\alpha-2 \choose k} \nonumber \\
&&\times 
\left[ \gamma \left(\frac{k+1}{2},w^2 \right) \right]_{w(\alpha,t)}^{w(\alpha,0)}
\end{eqnarray}  
where $K_m^{\alpha} = e^{1/\alpha} \, \alpha^{3/2 - m - 2 \, \alpha}$,
$w(\alpha, t) = \sqrt{\alpha} (e^{-t} -
1/\alpha)$. 
The infinite series only converges in the interval $\alpha \in [0,2)$, but it
becomes finite for half-integer and integer values $\alpha = n/2$ with
$n=0,1,2,\ldots$.  
Using the identities $\gamma (a+1,x) = a \gamma (a,x) - x^a e^{-x}$, $\gamma
(1, x) = 1 - e^{-x}$ and 
$\gamma (1/2,x) = \sqrt{\pi} \, \text{erf} (\sqrt{x})$~\cite{abramowitz+stegun}, 
we find, for example, $P(1;t)$ for
$\epsilon = -1/2$ to be
\begin{eqnarray}
P(1;t) = 1 + 2 \, \sqrt{e} \left[ \sqrt{\frac{\pi}{2}} \, \text{erf}(w) - (2 +
  \sqrt{2}\,w) \, e^{-w^2} \right]_{w(1/2,t)}^{w(1/2,0)} \label{eq:one-half-sol}
\end{eqnarray}
and for $\epsilon = -1$ to be
\begin{eqnarray}
P(1;t) = 1 + \left[ \sqrt{\frac{\pi}{4}} \, \text{erf}(w) - (2 +
  3 \,w + 2\,w^2) \, e^{-w^2} \right]_{w(1,t)}^{w(1,0)} \label{eq:one-sol}
\end{eqnarray}
where $\text{erf}(x)$ denotes the error function.
%~\cite{abramowitz+stegun}.  
It follows from Eqs.~(\ref{eq:one-half-sol}) and (\ref{eq:one-sol}) that 
the time-dependent coverage, $\theta(t)=1-P(1;t)$, asymptotically
approaches the critical value $\theta_J \simeq 0.876681$ as
\begin{eqnarray}
  \theta(t) &=& \sqrt{2 \pi e} \,
  \left(\text{erf}(-\sqrt{2})-\text{erf}(-1/\sqrt{2})\right)+ 2 \nonumber \\
  &-& %4  e^{-\frac{3}{2}} \exp(-2\,t) 
  4 \, e^{-\frac{3}{2}-2t} + \mathrm{O}\,((w(1/2,t)-w_\infty)^3)
  \label{eq:one-half-approach-to-jamming}
\end{eqnarray}
and $\theta_J \simeq 0.885296$ as 
\begin{eqnarray}
  \theta(t) &=& e^{-1} -2 + \text{erf}(-1)/2 +  %2 e^{-1} \exp(-3\,t) 
  2 \, e^{-1-3t} \nonumber \\
  && + \mathrm{O}\,((w(1,t)-w_\infty(1))^4)
%\sqrt{2 \pi e} \,
%  \left(\text{erf}(-\sqrt{2})-\text{erf}(-1/\sqrt{2})\right)+ 2 \nonumber \\
%&-& 2  e^{-\frac{3}{2}}
%  \exp(-2\,t) + \mathrm{O}\,((w(1/2,t)-w_\infty)^3)~ 
  \label{eq:one-approach-to-jamming}
\end{eqnarray}
for $\epsilon = -1/2$ and $\epsilon = -1$, respectively.
These expressions have been 
obtained by expanding Eqs.~(\ref{eq:one-half-sol}) and (\ref{eq:one-sol})
about $w_\infty (\alpha) \equiv 
\lim_{t \to \infty} w(\alpha,t) = -1/\sqrt{\alpha}$ such that $w - w_\infty =
\sqrt{\alpha} e^{-t}$.
The time-dependent coverage for any  $\epsilon=-n/2$ can be obtained
analytically  in a similar manner.
The important point
to note here is that Eqs.~(\ref{eq:one-half-approach-to-jamming}) and
(\ref{eq:one-approach-to-jamming}) suggest that for a given $n$, all terms of 
order smaller or equal $n$ will drop out of the expansion so that the leading
time-dependent term in the approach to jamming when $t \to \infty$ is $\exp(-t/
\tau)$ with $\tau = 1/(n+1)$.  
Extrapolating to any value for $\epsilon$, the characteristic time $\tau$ to
jamming for the one-dimensional process is then given by $\tau =
1/(1-2\epsilon)$ which can easily be verified numerically.
This form of the temporal approach to jamming also follows from
the time dependence of the sticking probability $P(2;t)$ given by
Eq.~(\ref{eq:1d_csa_n_sol}).

\begin{figure}
  \begin{center}    
    \scalebox{0.30}{\includegraphics[angle=0]{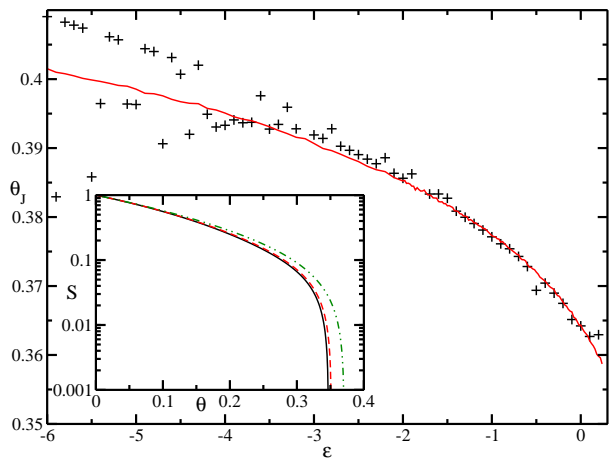}}
  \end{center}
  \caption{(color online).
    Comparison of jamming coverage $\theta_J$ for $\epsilon = -5.0,
    \ldots, 0.24$ from the series expansion up to order $13$ using
    Pad\'e approximant $[6,6]$, $[6,7]$ and $[7,6]$ for the optimization
    described in the text ($+$) and from MC simulations for
    a $200\times200$ - lattice and $100$ iterations (red $\sline$).
    Errors in
    simulation data points are of order $0.001$.
    Inset: Sticking probability $S$ as function of coverage $\theta$ up to
    order $N = 9$ for three different $ \epsilon = 0.1 (\sline)$, $\epsilon
    = 0.0 (\dashline)$, and
    $\epsilon = -0.5$ (\ddashline).
  } 
  \label{fig:jamcov-series-sim}
\end{figure}

The 2d case of monomer adsorption with NE and
NNN interaction cannot be handled analytically and the methods
of series expansion (SE) and Monte Carlo simulations (MC) were employed.
The SE is an Taylor expansion of any probability density $P(G;t)$ in $t$ using
the rate equations~(\ref{eq:CSAmaster}), $ P(G;t) = \sum_{n=0}
\frac{(-t)^n}{n!}\mathcal{L}^n P(G,t)$~\cite{evans_93, wang_00}.
For the coverage, the expansion of $P(\circ;t)$ is of interest.
We have implemented the algorithm introduced in~\cite{gan_96} with 
the extension
of allowing for polynomial values in $\epsilon$ in order to compute
the coefficients of the series numerically.
The series coefficients up to order $13$ were computed (see
Tab.~\ref{tab:coeffs}) using this algorithm. 
With these coefficients and the first terms in the expansion of 
$
P \big(
  \begin{array}{c@{\hspace{1pt}}c@{\hspace{1pt}}c}
    \vspace{-7.5pt} & \circ &     \\
    \vspace{-7.5pt} \circ & \circ & \circ\\
    & \circ & 
  \end{array}; t) \equiv P(G_1;t)
$,
we can also calculate the expansion of the
sticking probability,  
$S(\theta) = P(G_1, \theta)$,
as a function of the coverage $\theta (t)$.
The first few terms are
\begin{eqnarray}
  S(\theta) &=& 1 - 5 \, \theta + (6 - 4 \epsilon) \, \theta^2 +
  \frac{8}{3}(1 + 3 
  \epsilon - 5 \epsilon^2) \, \theta^3 \nonumber \\ 
  &&- \frac{2}{3} (1  - 26 \epsilon - 13
  \epsilon^2  + 84 \epsilon^3) \, \theta^4 
  + \mathrm{O}(\theta^5) \label{eq:S}
\end{eqnarray}
The interaction hardly affects the sticking probability for small
coverages (see
inset in Fig.~\ref{fig:jamcov-series-sim}) and the effect can be seen
only close to the jamming limit. 
This is what one would expect, given the short-range nature of the
interaction.
The value of the sticking probability for attractive (repulsive) 
interaction are
expectedly greater (smaller) than for the adsorption without NNN
interaction.

%%%%%%%%%%%%%%%%%%%%%%%%%%%%%%%%%%%%%%%%%%%%%%%%%%

\begin{table}
  \begin{center}
    \caption{The series coefficients $c_{nm}$ for the Taylor expansion
      $P(\circ;t) = 
      \sum_{n=0}^N \sum_{m=0}^{n-1} \, (-t)^n \, \epsilon^m \, c_{nm}/ n!$ up to order
      $N = 13$. \label{tab:coeffs}}
    \begin{ruledtabular}
      \begin{tabular}{rrrrrrrrr}
        $n$ & $m$ & $c_{nm}$ & $n$ & $m$ & $c_{nm}$ & $n$ & $m$ & $c_{nm}$ \\
        \hline
        0 & 0 &  1 &  & 6 & -1408 &  & 3 &  2351030011040 \\
        &  &  &  &  &  &  & 4 &  1700297848328 \\
        1 & 0 &  1 & 8 & 0 & 12017245 &  & 5 &  608141922992 \\
        &  &  &  & 1 & 58696340 &  & 6 &  93227257416 \\
        2 & 0 &  5 &  & 2 & 102985272 &  & 7 &  4385436272 \\
        & 1 &  4 &  & 3 & 75819336 &  & 8 &  -115745776 \\
        &  &  &  & 4 & 21447496 &  & 9 &  44930880 \\
        3 & 0 &  37 &  & 5 & 1699920 &  & 10 &  -6324224 \\
        & 1 &  56 &  & 6 & -42368 &  &  &  \\
        & 2 &  8 &  & 7 & 11008  & 12 & 0 &  1903886785277 \\
        &  &  &  &  &  &  & 1 &  14838958395140 \\
        4 & 0 &  349 & 9 & 0 &  213321717 &  & 2 &  46840094488488 \\
        & 1 &  756 &  & 1 &  1194111320 &  & 3 &  75901023988384 \\
        & 2 &  328 &  & 2 &  2505972296 &  & 4 &  66977890612768 \\
        & 3 &  16 &  & 3 &  2378270528 &  & 5 &  31546273918392 \\
        &  &  &  & 4 &  1002645360 &  & 6 &  7295131690264 \\
        5 & 0 &  3925 &  & 5 &  157668392 &  & 7 &  675665248656 \\
        & 1 &  11080 &  & 6 &  4282992 &  & 8 &  12950433024 \\
        & 2 &  8344 &  & 7 &  441776 &  & 9 &  1365586784 \\
        & 3 &  1448 &  & 8 &  -90240 &  & 10 &  -436513376 \\
        & 4 &  -32 &  &  &  &  & 11 &  52077696 \\
        &  &  & 10 & 0 &  4113044061 &  &  &  \\
        6 & 0 &  50845 &  & 1 &  25999942820 & 13 & 0 &  45187885535477 \\
        & 1 &  177716 &  & 2 &  63624902448 &  & 1 &  385950877646856 \\
        & 2 &  192112 &  & 3 &  74332714408 &  & 2 &  1357386708834952 \\
        & 3 &  65712 &  & 4 &  42302559256 &  & 3 &  2513522034088536 \\
        & 4 &  3232 &  & 5 &  10661184336 &  & 4 &  2628475342601104 \\
        & 5 &  192 &  & 6 &  903506912 &  & 5 &  1550899615251504 \\
        &  &  &  & 7 &  21751616 &  & 6 &  491117645970296 \\
        7 & 0 & 742165 &  & 8 &  -4520816 &  & 7 &  73659458920040 \\
        & 1 & 3104424 &  & 9 &  755712 &  & 8 &  4066062738848 \\
        & 2 & 4393304 &  &  &  &  & 9 &  53580650752 \\
        & 3 & 2339128 & 11 & 0 &  85493084853 &  & 10 &  -14364063968 \\
        & 4 & 365928 &  & 1 &  603053910056 &  & 11 &  4113892304 \\
        & 5 & 11472 &  & 2 &  1688460211624 &  & 12 &  -416352000 \\
      \end{tabular}
    \end{ruledtabular}
  \end{center}
\end{table}

%%%%%%%%%%%%%%%%%%%%%%%%%%%%%%%%%%%%%%%%%%%%%%%%%%%

To obtain estimates for $\theta_J$ from the series  
the standard Pad\'e approximants~\cite{dickman_91_RSA, baker_96}
$[n,m]$ (where $n$ and $m$ are the orders of the polynomials in the
numerator and denominator of the Pad\'e approximant) 
were  used. 
First, the transformation of variables, $y = (1-\exp(-(1-b \epsilon)t))/(1-b
\epsilon)$ (similar to that used in Ref.~\cite{de_oliveira_92}), 
was carried out 
with $b$ being an adjustable parameter. 
It is clear from the preceeding discussion that this mimics the approach to
jamming in 1d where $b$ take the value $b=2$.
It has been found before~\cite{dickman_91_RSA} that using the knowledge of the
temporal behavior of the coverage in 1d for the transformation of variables can
considerably improve estimates in 2d.
However, instead of choosing a value for $b$, we use it to make the
estimates of  $\theta_J$ independent of the choice of Pad\'e approximant.
In order to do so, the free parameter $b$ has been found
by minimizing the cost function 
$C(\epsilon, b) = \big(\theta_J (\epsilon, [6,6],b) - \theta_J
(\epsilon,[6,7],b) \big)^2 + \big(\theta_J (\epsilon, [6,6],b) - \theta_J
(\epsilon, [7,6],b)\big)^2 + \big(\theta_J (\epsilon, [6,7],b) - \theta_J
(\epsilon, [7,6], b)\big)^2$ with respect to $b$  
where  $\theta_J (\epsilon,[n,m],b)$ is the jamming coverages 
obtained for the highest-order Pad\'e approximants available.  

%%%%%%%%%%%%%%%%%%%%%%%%%%%%%%%%%%%%%%%%%%%%%%%%%%%%%%%%%%%%%%%%%%%%%%%%%%%%%%%
%%%%%%%%%%%%%%%%%%%%%%%%%%%%%%%%%%%%%%%%%%%%%%%%%%%%%%%%%%%%%%%%%%%%%%%%%%%%%%%
\begin{figure}
  \begin{center}
    \scalebox{0.30}{\includegraphics[angle=0]{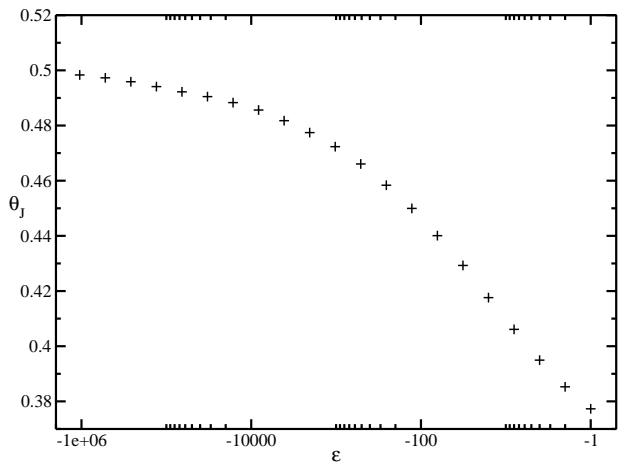}}
  \end{center}
  \caption{(color online).
    Jamming coverage $\theta_J$ for $\epsilon = -1.0, \ldots, -1.0
    \times 10^6$ from MC simulations for a $200\times200$ - lattice and $100$ iterations.
    Errors in
    simulation data points are of order $0.001$ and thus smaller than
    the symbol size.
  } 
  \label{fig:large-neg-eps-MC}
\end{figure}
%%%%%%%%%%%%%%%%%%%%%%%%%%%%%%%%%%%%%%%%%%%%%%%%%%%%%%%%%%%%%%%%%%%%%%%%%%%%%%%

The results of this analysis are presented in
Fig.~\ref{fig:jamcov-series-sim} and compared with the values of
jamming coverage calculated numerically by MC simulations. 
In the MC simulations,   
an event-driven algorithm~\cite{gillespie_92, evans_93, wang_00} was
used. 
Within this algorithm, all the susceptible binding sites were grouped
depending on the number of occupied NNN sites. 
A binding site for the next adsorption event is then drawn randomly
out of a group according to the rates $r_n$ and the waiting times are 
distributed exponentially.  
The results for the jamming coverage calculated numerically for a wide
range of interactions are presented in
Figs.~\ref{fig:jamcov-series-sim} and \ref{fig:large-neg-eps-MC}.

In Fig.~\ref{fig:jamcov-series-sim}, one can clearly see that the series
expansion and the MC simulations 
agree for $\epsilon \in [-2.0, 0.1]$.
The relative effect of the interaction is strongest around the point $\epsilon
= 0$ and then flattens as $\epsilon$ becomes more negative, i.e.\ the
interaction becomes more attractive.
In Fig.~\ref{fig:large-neg-eps-MC}, we can see that only for very large negative $\epsilon$,
$\epsilon \alt -10^4$, the jamming coverage comes within a few percent of the
ideal coverage of $\theta_J = 0.5$.

In conclusion, we have presented the analysis of a cooperative
sequential adsorption model, that takes into
account the effects of nearest-neighbor exclusion as well as 
physically important (repulsive or attractive) interactions between
next-nearest neighbours.    
A one- and two-dimensional process, dimer adsorption and
monomer adsorption with nearest-neighbor exclusion, respectively, have
been studied. 
For the one-dimensional process, we computed the coverage analytically for 
for a family of special cases, where the interaction parameter
$\epsilon$ takes negative half- and integer values.
This allowed us to compute the jamming coverage and
to extract the temporal approach to jamming.
For the two-dimensional process, we have computed the series expansion for the
coverage as a function of time and have found the jamming coverage for
various strengths of interaction. 
Monte Carlo simulations convincingly support the  
series expansion results and provide estimates for the jamming
coverage that are unaccessible to the series expansion, thus
demonstrating a slow convergence to the ideal coverage 
with increasing attractive interaction.

CJN would like to acknowledge the UK EPSRC and the Cambridge European
Trust for financial support.

\end{document}